\documentclass[pre,reprint,floats]{revtex4-1}

\newcommand{\rmd}{\ensuremath{\mathrm{d}}}
\newcommand{\rmi}{\ensuremath{\mathrm{i}}}

\newcommand{\erf}{\ensuremath{\mathrm{erf}}}

\usepackage[lofdepth,lotdepth,caption=false]{subfig}
\usepackage{amsmath}
\usepackage{amssymb}
\usepackage{mathtools}
\usepackage{graphicx}
\usepackage{dcolumn}
\usepackage{color}
\usepackage{blindtext}
\usepackage[outdir=./figs]{epstopdf}
\usepackage{float}
\begin{document}

%\title{Kinetics of charge carriers at the negatively biased plasma-solid interface}
\title{Charge kinetics across a negatively biased semiconducting plasma-solid interface}

\author{K. Rasek, F. X. Bronold and H. Fehske}
\date{\today}

 \affiliation{Institut f{\"u}r Physik,
	Universit{\"a}t Greifswald, 17489 Greifswald, Germany }

\begin{abstract}
An investigation of the selfconsistent ambipolar charge kinetics across a negatively
biased semiconducting plasma-solid interface is presented. For the specific 
case of a thin germanium layer with nonpolar electron-phonon scattering, sandwiched 
between an Ohmic contact and a collisionless argon plasma, we calculate the current-voltage 
characteristic and show that it is affected by the electron microphysics of the 
semiconductor. We also obtain the spatially and energetically resolved fluxes and charge 
distributions inside the layer, visualizing thereby the behavior of the charge carriers 
responsible for the charge transport. Albeit not quantitative, because of the crude model 
for the germanium band structure and the neglect of particle-nonconserving scattering 
processes, such as impact ionization and electron-hole recombination, which at the energies 
involved cannot be neglected, our results clearly indicate (i) the current through the 
interface is carried by rather hot carriers and (ii) the perfect absorber model, often used 
for the description of charge transport across plasma-solid interfaces, cannot be maintained 
for semiconducting interfaces. 
\end{abstract}

\maketitle

\section{Introduction}

Low-temperature gas discharges are bound by solid objects, acting either as confining walls 
or as electrodes. To maintain the discharge, an electric current has to flow across the 
plasma-electrode interface. Since the plasma of the discharge contains electrons, ions, and 
radicals, the charge transfer across the interface is ambipolar, consisting of electrons deposited 
into the electrode and extracted electrons (holes) arising from the neutralization of ions and the 
de-excitation of radicals. Hence, inside the electrode, a flux of electrons and holes builds up 
whose fate depends on the electron microphysics of the electrode material. 

The transport scenario just described is in some sense obvious, hardly addressed in textbooks
on plasma discharges~\cite{Franklin76,LL05}, and of course qualitatively known since the 
beginning of gaseous electronics~\cite{LM24}. On a fundamental level, however, it implies a 
subtle interplay of gaseous and solid state transport processes in any man-made gas discharge. 
Its investigation may thus perhaps bear novel possibilities for controlling discharges by 
manipulating processes inside the electrodes. The electric breakdown in dielectric barrier 
discharges, for instance, depends on the charge distribution inside the 
dielectric~\cite{NTH18,PRS16,WYB05,MSG03}, and hence on the transport processes to which they 
give rise to. Revealing how gaseous and solid-based charge transport merge at the plasma-solid 
interface may thus allow to control the breakdown by a judicious choice of the dielectric. It 
may also suggest optimization strategies for large scale industrial barrier 
discharges~\cite{Kogelschatz03}. However, a quantitative description of charge transport across 
the interface will be most beneficial for the further development of microdischarges embedded 
in semiconducting substrates~\cite{CMS19,EPC13,TWH11,DOL10}, where the time and length scales 
of electron transport and energy relaxation are no longer well separated (see the Introduction
of Ref.~\cite{BF17} for a discussion of this point). It is thus the purpose of this work to 
provide first steps towards a selfconsistent kinetic description of the ambipolar charge 
transport across a biased semiconducting plasma-solid interface.

In an attempt to model the whole electric double layer forming at a plasma-solid 
interface, and not only the positive space charge on the plasma side (that is, the plasma
sheath~\cite{SB90,Riemann91,Franklin03,Brinkmann09,Robertson13}), we recently developed a model for 
floating dielectric interfaces that treats the electrons and holes in the solid on the same kinetic 
footing as the electrons and ions in the plasma~\cite{RBF20,BF17}. The model links the electron-ion 
plasma to the electron-hole plasma inside the dielectric by allowing electrons to cross the interface 
in both directions. Electrons from the plasma may thus not only enter the conduction band of the solid 
by traversing the surface potential but also leave it due to internal backscattering and subsequent 
traversal of the surface potential in the reverse direction. In addition, electrons can be 
extracted from the valence band by the neutralization of ions. Tracking the charge distributions
by a set of Boltzmann equations, the resulting charge imbalance can be determined and used as a 
source in the Poisson equation, determining the selfconsistent electric field, which in turn 
influences the kinetics of the charge carriers on both sides of the interface. 
The selfconsistent solution of the Boltzmann-Poisson system for the distribution functions and the
electric potential is thus at the core of our approach. For the floating interface, where no net 
flux is flowing through the interface, we have solved this set of equations under simplifying 
assumptions~\cite{RBF20}. We now extend the model to an interface which carries a net current and 
remedy also some of the limiting simplifications used before. 
	
As a first step towards a realistic treatment of the current-carrying interface, we consider a 
planar semiconductor of finite thickness, in which electrons and holes loose or gain energy by 
scattering on optical phonons, sandwiched between an Ohmic contact and a collisionless plasma.
At the interface between the plasma and the semiconductor, electrons may be reflected when impinging 
on the interface from either side, while ions are neutralized with unit probability. At the other 
end of the semiconductor, the Ohmic metal serves as a sink for electrons and holes. The resulting setup
resembles thus a Langmuir probe~\cite{ML26,Cherrington82,Lam65} coated with a semiconducting layer. A 
negatively biased probe attracts fluxes of electrons and ions, the magnitude of which depends on the
bias voltage. At large enough negative bias, electrons cannot reach the probe anymore, resulting in 
a negative net flux due solely to ions. Otherwise the flux is dominated by the electrons, due to 
their lower mass and higher temperature. By calculating the functional dependence of the net flux 
on the bias voltage, that is, the current-voltage characteristic, and comparing it with the 
characteristic of a perfectly absorbing interface, which assumes that any charge carriers hitting 
the interface get instantaneously absorbed and never enter the plasma again, we can determine 
the influence the electron microphysics of the semiconducting layer has on the electric current flowing 
through the device. Our simulations show that the assumptions of the perfect absorber model cannot 
be maintained. The charge kinetics inside the semiconductor is an essential part of the kinetics of 
the gas discharge and should thus be included in its modeling.

The outline of the paper is as follows. In the two parts of Sec.~\ref{sec:Theory} we present the 
equations of the kinetic model and discuss its numerical implementation and solution, focusing on 
aspects which differ from our treatment of the floating interface. Using for illustration 
germanium in contact with an argon plasma, numerical results 
are given in Sec.~\ref{sec:results}. It is divided into three parts, corresponding to the three 
perspectives from which one may consider the interface. First, in subsection~\ref{sec:BP}, it is regarded as 
an electric device, discussing thus the current-voltage characteristic that results from the kinetic 
theory. Spatially resolved macroscopic properties, such as density distributions and potential profiles, 
are discussed in subsection~\ref{sec:MacP}, while a microscopic view, based on spatially and energetically 
resolved distribution functions is presented in subsection~\ref{sec:MicP}. The paper concludes in 
Sec.~\ref{sec:conclusion} with an outlook to what could be the next steps.
	
\section{Theoretical framework}\label{sec:Theory}

\begin{figure}
\includegraphics[width=\linewidth]{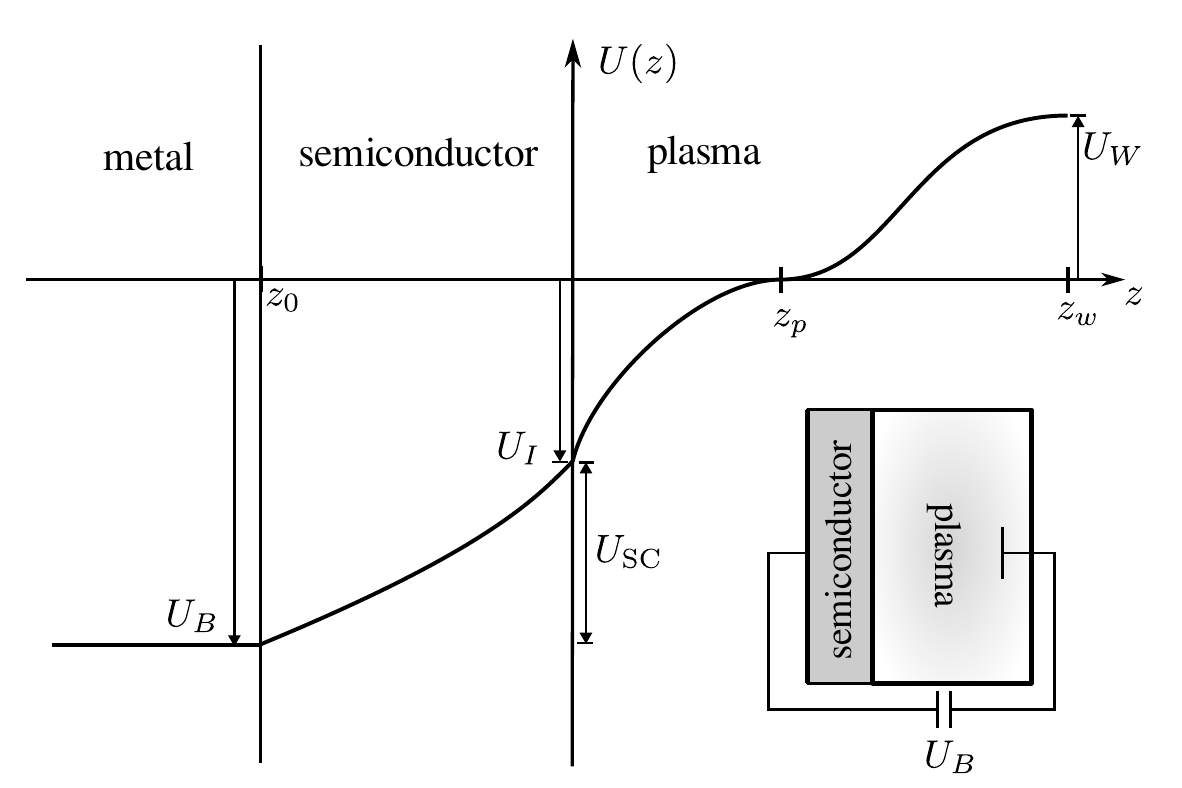}
\caption{Illustration of the electric potential across a negatively biased semiconducting plasma-solid 
interface (not to scale). The bias voltage $U_B$ is applied between the bulk plasma and the Ohmic 
contact to the left of the semiconducting layer, as sketched in the bottom right. Measuring the potential 
with respect to its value in the plasma bulk, $U_B=U_I-U_{SC}$, with $U_I$ the sheath (interface) 
potential at $z=0$ and $U_{SC}$ the potential drop across the semiconducting layer. In the 
Schwager-Birdsall approach~\cite{SB90} to the collisionless plasma, the presheath potential $U_W$ 
stretches over the entire region $z>z_p$, where $z_p$ is the location of the bulk plasma, which is
effectively infinitely far away from the interface.}
\label{fig:concept}
\end{figure}

\subsection{Model equations}\label{sec:Eqs}
%As in our previous work addressing the floating interface~\cite{RBF20}, 
The basis for the kinetic modeling 
of the current-carrying interface is a set of Boltzmann equations, describing the dynamics of the 
distribution functions of the charge carriers on both sides of the interface, augmented by the Poisson
equation for the electric potential and matching as well as boundary conditions for the distribution 
functions and the potential~\cite{BF17}. Due to the particle flux through the interface, the boundary
conditions differ from the ones used for a floating interface. To make the modeling more realistic, we
improve in this work also the matching conditions at the plasma-semiconductor interface, considering now 
realistic injection energies for holes and the possibility for electrons to be quantum-mechanically 
reflected. If not noted otherwise, all equations are written in atomic units, measuring energy in 
Rydbergs, length in Bohr radii, and masses in electron masses. 
	
Figure~\ref{fig:concept} shows the electric potential energy $U(z)$ across the interface in the manner 
it is implemented in the model of this paper. Within the plasma ($z>0$), we employ the Schwager-Birdsall 
approach~\cite{SB90} for a collisionless plasma to model the merging of the plasma sheath with the bulk 
plasma. The sheath potential $U_I=U(0)$ is thus the potential difference between the interface at 
$z=0$ and a point $z_p$, where the bulk plasma is established and which also serves as the reference
point from which electric potentials are measured. The presheath potential $U_W=U(z_w)$ accelerates 
the ions to make the Bohm criterion satisfiable at $z_p$. Since $z_p$ tends to infinity, the presheath 
does not belong to the physically relevant part of the plasma-solid interface. In the following, we adopt 
the sign convention that $U_B=U(z_0)$ and $U_I$ are negative, whereas $U_W$ and $U_{SC}$ are positive. 
Hence, 
\begin{align}
U_B=U_I-U_{SC}~. 
\label{VoltageDrop}
\end{align}
	
When no collisions are considered within the plasma, the particle densities can readily be expressed as 
functions of $U(z)$. It is thus practical to use the once integrated Poisson equation to calculate the 
electric field in the form
\begin{equation}
	\label{eq:U'P}
	\frac{\rmd U}{\rmd z} = \mathcal{E}(z) =
	\left(16 \pi\int_{U(z_p)}^{U(z)}\rmd U' n(U')\right)^{1/2} ~,
\end{equation}
with $U(z_p) = 0$ and also $\mathcal{E}(z_p) = 0$ since the bulk plasma is field-free.
	
The semiconducting material, having a dielectric constant $\varepsilon$, stretches from $z=z_0<0$ to $z=0$. 
It has thus a fixed width and it is more reasonable to keep $z$ as the spatial variable. Hence,  
\begin{equation}
\label{eq:U'SC}
\mathcal{E}(0^-)-\mathcal{E}(z) = 
\frac{8\pi}{\varepsilon}\int_{z}^{0}\rmd z' n(z') ~.
\end{equation}
Using the matching condition for the electric potential at the interface,
\begin{equation}
	\label{eq:matching}
	\varepsilon \mathcal{E} (0^-) = \mathcal{E}(0^+)~,
\end{equation}
the magnitude of the electric field, and thus the electric potential, across the entire interface can be 
determined for a given charge density
\begin{multline}
\label{eq:ntot}
n(z) = \left[n_e(z) -  n_i(z)\right]\Theta(z)\\
+ \left[n_*(z) - n_h(z) + n_A -n_D \right] \Theta(-z)~,
\end{multline}
where $n_s(z)$ denotes the density of electrons $(s=e)$, ions $(s=i)$, conduction band electrons $(s=*)$, 
and valence band holes $(s=h)$. For an intrinsic semiconductor, the acceptor ($n_A$) and donor densities ($n_D$)
are absent. Inside the metal, the electric potential is constant and the electric field vanishes.

The Boltzmann equations governing the distribution functions $F_s^\gtrless(z,E,T)$ for the charge carriers 
on either side of the interface are given by~\cite{RBF20} 
\begin{equation}
\label{eq:BEQ}
	\pm v_s \frac{\partial}{\partial z} F^\gtrless_s = 
	\Phi^\gtrless_s - \gamma^\gtrless_s F^\gtrless_s~,
\end{equation}
with $\Phi^\gtrless_s$ the in-scattering part of the collision integral, $\gamma^\gtrless_s$ the scattering 
rate, both will be specified below, and 
\begin{equation}
	\label{eq:vs}
	v_s(z,E,T) = 2\sqrt{m_s^{-1}(E - U_s(z)-T)}~
\end{equation}
the modulus of the velocity in $z-$direction. 

For brevity, the independent variables, which are the total energy $E$, the lateral kinetic energy 
$T$, and the spatial variable $z$ are suppressed in Eq.~\eqref{eq:BEQ} and the distribution functions 
for left- and right-moving particles are distinguished by the superscripts $<$ and $>$. In the lateral
directions the interface is isotropic. Equation~\eqref{eq:vs} holds only for parabolic dispersions,
which are of course valid for the free charge carriers on the plasma side of the interface, but for 
the free carriers inside the solid it is an approximation specified by effective masses. Anticipating
to use for illustration germanium, we take for the effective electron mass 
the density-of-state effective mass, $m_* = \left(9 m_lm_t^2\right)^{\frac{1}{3}}$, and for the 
hole mass the average of the masses of light and heavy holes. With the numerical values from 
Ref.~\cite{JR83}, we then obtain the masses given in Table~\ref{tab:parameters}. However, since the 
injection energies for electrons and holes are rather high, the parabolic dispersion is only a crude
approximation to the band structure. More than one valley as well as nonparabolicities should be 
actually considered. But it is beyond the scope of this exploratory work. The species potential 
$U_s(z)$, finally, takes each species' charge and energy offset into account, relating thus to the 
electric potential $U(z)$ via $U_i = U$, $U_e  = -U$, $U_* = -U-\chi$, and $U_h = U + E_g + \chi$, 
with the electron affinity $\chi$ and band gap $E_g$.

\begin{table}
\setlength\extrarowheight{3pt}
\caption{Material parameters for the germanium layer~\cite{JR83} and the argon plasma facing it.
}
\label{tab:parameters}
\begin{ruledtabular}
	\begin{tabular}{llll}
$k_B T_{SC} (\text{eV})$    &	0.025	  &   $m_{*}(m_e)$	                 &  0.2   \\
$k_B T_{i} (\text{eV})$	    &	0.025  	  &   $m_h (m_e)$	                 &  0.34  \\
$k_B T_{e} (\text{eV})$	    &	2         &   $m_i(m_e)$	                 &  73551 \\
$D_tK (10^8\textrm{eV/cm})$ &	9.5	  &   $n_i (10^{13}\textrm{cm}^{-3})$    &  2	  \\
$n_p (\textrm{cm}^{-3})$    &	$10^{13}$ &   $\rho(\textrm{g/cm}^{3})$	         &  5.32  \\
$\varepsilon$	   	    &	16.2	  &   $\hbar\omega_0(\text{eV})$         &  0.037 \\
$E_g(\text{eV})$	    &	0.67	  &   $\chi (\textrm{eV})$	         &  4	  \\
$I_\textrm{inj}(\text{eV})$ &	15.76	  &   $\Gamma_\mathrm{inj} (\textrm{eV})$ &  3	
		\end{tabular}
	\end{ruledtabular}
\end{table}

While no collisions are considered on the plasma side, implying $\Phi_{e,i} = \gamma_{e,i} = 0$, 
within the semiconductor, we include collisions with optical phonons. For nonpolar materials, such as
silicon or germanium, the phonon collision integral is isotropic. Hence, no distinction between 
left- and right-moving particles must be taken into account, implying $\Phi_s^>=\Phi_s^<=\Phi_s$ 
and $\gamma_s^>=\gamma_s^<=\gamma_s$. The collision rates entering the 
Boltzmann equations for conduction band electrons and valence band holes are then~\cite{Ridley99,Roth92}
\begin{multline}\label{eq:gamma}
\gamma_s(z,E) = \frac{(D_tK)^2m_s}{4 \pi\rho \omega_0}\sqrt{m_s} \left [ n_b \sqrt{E+\hbar\omega_0 -U_s}\right. \\
\left.+ (n_b+1)\sqrt{E-\hbar\omega_0 -U_s}\right]~,
\end{multline}
where the second term in the square brackets only applies if the argument of the root is positive, while 
the in-scattering parts of the collision integrals read 
\begin{multline}
\Phi_s(z,E) = \frac{(D_tK)^2}{8 \pi\rho \omega_0} \left[ n_b N_s(z,E-\hbar\omega_0)\right. \\
\left.	+ (n_b+1)N_s(z,E+\hbar\omega_0)\right]~,
\end{multline}	
with the optical deformation potential $D_t K$, the mass density $\rho$, the optical phonon frequency 
$\omega_0$, the phonon occupation number $n_b = 1/\left(\exp(\hbar\omega_0/k_BT_{SC})-1\right)$, and
\begin{equation}
\label{eq:Is}
N_s(z,E) = m_s\sum_{\gtrless}\int_{0}^{E-U_s(z)}\rmd T\frac{F_s^\gtrless(z,E,T)}{v_s(z,E,T)}~,
\end{equation}
the spatially and energetically resolved density of the species $s$, from which the densities entering
the Poisson equation~\eqref{eq:U'SC} follow by one more integration,
\begin{equation}
\label{eq:n}
n_s(z) = \int_{U_s(z)}^\infty \frac{\mathrm{d}E}{8\pi^2}N_s(z,E)~.
\end{equation}
For germanium, the material parameters required for $\gamma_{*,h}$ and $\Phi_{*,h}$ are given in 
Table~\ref{tab:parameters}. 
	
As in our previous work~\cite{RBF20}, we solve the equations on the plasma side analytically and use the 
iterative approach by Grinberg and Luryi~\cite{GL92} inside the semiconductor. The boundary conditions at 
the outer limits of the kinetically modeled interval are essential to determine the solution. On the right 
boundary, at $z=z_w$, the Schwager-Birdsall model~\cite{SB90} prescribes half Maxwellian distributions
\begin{equation}
\label{eq:LM}
F_s^\mathrm{LM}(z_w)=
n_s^\mathrm{LM}\left( \frac{4\pi}{k_B T_s m_s}\right)^{3/2} \exp\left( -\frac{E-U_s(z_w)}{k_BT_s} \right)~,
\end{equation}
with densities $n_{i,e}^\textrm{LM}$ such that at $z=z_p$ the densities for both electrons and ions are 
equal to the plasma density $n_p$. On the left boundary, at $z=z_0$, that is, at the metal-semiconductor  
interface, the boundary condition is not strictly known. Following standard semiconductor device 
modeling~\cite{Moglestue93}, we assume an Ohmic contact, implying at $z=z_0$ Maxwellian distributions 
with densities and temperatures applying to the bulk of the semiconductor. For an undoped, intrinsic semiconductor, 
for instance, the boundary condition at $z=z_0$ is thus given by~\eqref{eq:LM} with $z_w$ replaced by $z_0$, 
$T_s=T_{SC}$, and $n_s^\mathrm{LM}=n_{\rm int}$, where $n_{\rm int}$ is the intrinsic density of the 
semiconductor. Put together, we thus have to enforce at the system boundaries,
\begin{subequations}
\begin{align}
F_s^> (z_0) = F_s^\mathrm{LM}(z_0)\hspace{30pt}\text{for }&s = h,*~, \label{eq:BC_z0}\\
F_s^< (z_w) = F_s^\mathrm{LM}(z_w)\hspace{30pt}\text{for }&s = i,e~.
\label{eq:BC_zw}
\end{align}
\end{subequations}
	
We also need to match distribution functions at the interface between the plasma and the semiconductor. For 
ions, we assume perfect neutralization, that is, an impinging ion extracts (injects) with unit probability
an electron (hole) from the valence band. For electrons, the interface is quantum-mechanically reflecting
with a reflection coefficient $R(E,T)$. Thus, the matching conditions at $z=0$ read 
\begin{subequations}	 
\label{eq:BCallfs}
\begin{align}
F_i^>(0,E,T) &= 0 \label{eq:BC_i}\\
F_h^<(0,E,T) &= F_h^>(0,E,T) + S_h^<(E,T)\label{eq:BC_h}\\
F_*^<(0,E,T')&= R(E,T)F_*^>(0,E,T')+ S_*^<(E,T')\label{eq:BC_*}\\
F_e^>(0,E,T) &= R(E,T)F_e^<(0,E,T) \nonumber\\
	&+ (1-R(E,T))F_*^>(0,E,T')\label{eq:BC_e}
\end{align}
\end{subequations}
\noindent with source terms $S^<_s$, describing injection of holes and electrons into the semiconductor, given by
\begin{subequations}	 
\begin{multline}
\label{eq:S_h}
S_h^<(E,T)= n_h^\mathrm{inj}\left( \frac{4\pi}{k_B T_h m_h}\right)^{\frac32} \\ 
\times \exp\left(-\frac{(E-U_h(0)-I_\mathrm{inj})^2}{\Gamma_\mathrm{inj}^2}\right)~,
\end{multline}
which is in fact independent of $T$, and 
\begin{equation}
\label{eq:S_*}
S_*^<(E,T')= (1-R(E,T))F_e^<(0,E,T)~.
\end{equation}
\end{subequations}
In the matching conditions for the electron distribution function, $T'=T/m_*$. The change 
in lateral energy from $T$ to $T'$ arises from the conservation of lateral momentum. Since, 
the effective mass $m_*<m_e$, the electron gains (looses) lateral energy while passing through 
the interface from the plasma (solid) side. The mass mismatch leads also to total reflection 
for electrons coming from the plasma when $E-U_*(0)<T^\prime$. 

\begin{figure}[t]
\includegraphics[width=\linewidth]{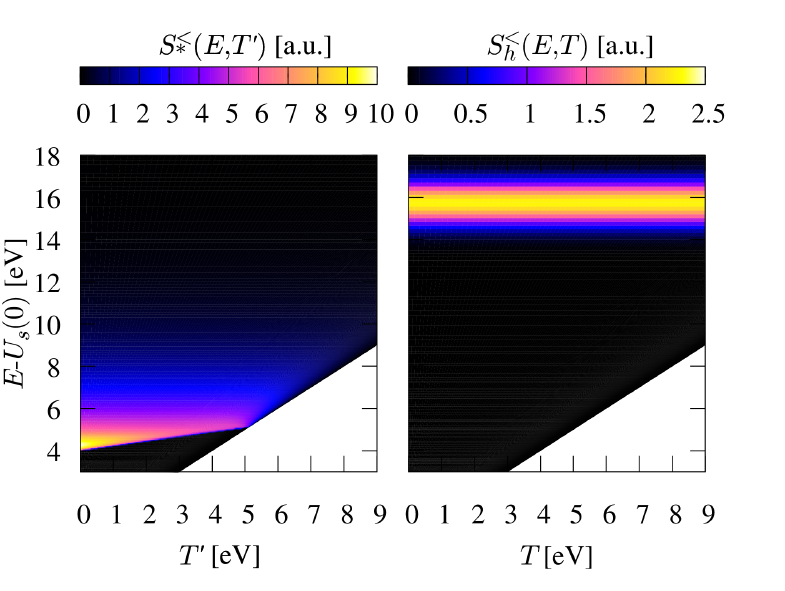}
\caption{(color online) Energy-resolved source functions $S_*^<$ (left) and $S_h^<$ (right) in 
arbitrary units for the floating interface, that is, the situation where electron and ion fluxes 
are equal. The material parameters are taken from Table~\ref{tab:parameters}, the thickness of the 
germanium layer is 1 $\mu\mathrm{m}$, $U_W=0.97\,\mathrm{eV}$, and $U_B=-5.125\,\mathrm{eV}$. Note, 
$T^\prime>E-U_*(0)$ and $T>E-U_h(0)$ are energetically not allowed. Since the injection of holes 
spreads over a larger energy range, the absolute values of $S_h^<$ are smaller than the values 
of $S_*^<$. Integrated over energy, however, the source functions ensure flux equality, as required 
for the floating interface.}
\label{fig:S_*}
\end{figure}

Both source terms are illustrated in Fig.~\ref{fig:S_*} for the parameters given in Table~\ref{tab:parameters},
which are used for the numerical calculations described in the next section. The normalization density 
$n_h^\textrm{inj}$ in the source term for the holes is chosen such that the flux is conserved across  
the interface. With this source term, holes are injected at the ionization energy $I_\textrm{inj}$ of an 
argon atom, homogeneously distributed in lateral direction. The width $\Gamma_\textrm{inj}$ accounts for 
an energy spread in the neutralization process. The source for electrons depends on the reflection 
coefficient $R(E,T)$. For a surface potential with the depth of the electron affinity $\chi$ and 
a $1/z$ tail on the plasma side due to the image charge (Schottky effect), the coefficient becomes, 
adapting results from Ref.~\cite{MacColl39}, 
\begin{equation}\label{eq:R}
R(E,T) =  \left|\frac{v_e(0,E,T)- y\,v_*(0,E,T')}{v_e(0,E,T)+ y^*\,v_*(0,E,T')}\right|^2~ 
\end{equation}
with 
\begin{equation}
y = -2\frac{W'_{\lambda,\frac{1}{2}}(\xi_0)}{W_{\lambda,\frac{1}{2}}(\xi_0)}~,
\end{equation}
where $W_{\lambda,\frac{1}{2}}(x)$ denotes the Whittaker function, $W'_{\lambda,\frac{1}{2}}(x)$ its first 
derivative with respect to $x$, 
\begin{align}
\lambda &= -\rmi \frac{\varepsilon -1}{\varepsilon+1}\frac{1}{\sqrt{8}\sqrt{E-T-U_I}}~,\\
\xi_0 &= \frac{\rmi \sqrt{2}}{\chi}\frac{\varepsilon -1}{\varepsilon+1} \sqrt{E-T-U_I}~,
\end{align}
and $y^*$ is the complex conjugate of $y$. As can be seen in Fig.~\ref{fig:S_*}, due to the high 
temperature, electrons are injected into the conduction band of the semiconductor over a wide range of 
energies, with most weight at the low energy cutoff given by $E-U_*(0)=\chi$.  
	
Since the plasma is treated collisionless, the solutions of the Boltzmann equations on the plasma side
are completely determined by the profile of the electric potential and the boundary conditions at $z=0$ 
and $z=z_w$. The latter are given by~\eqref{eq:BC_zw}, where the ions are restricted to energies above 
the presheath potential $U_W$, which needs to be determined selfconsistently. It is responsible for the 
acceleration of ions before reaching the sheath and can be determined from the generalized Bohm  
criterion~\cite{Riemann91},
\begin{align}\label{eq:GBC}
\partial_U(n_e-n_i)|_{z_p}\geq 0~.
\end{align}
In our formalism, it follows from Eq.~\eqref{eq:U'P}, where the radicand needs to be positive for 
$z \lesssim z_p$. As usual, we enforce marginal fulfillment. Thus, $U_W$ is determined by Eq.~\eqref{eq:GBC}
with the equal sign. In this work, we will prescribe the plasma density $n_p$. Enforcing Eq.~\eqref{eq:GBC} 
instead of $\mathcal{E}(z_w)=0$, as in~\cite{RBF20}, is then numerically advantageous. 

For the current-voltage characteristic we need the particle fluxes. As for the densities given in
Eq.~\eqref{eq:Is}, we initially define energy-resolved fluxes,
\begin{equation}
\label{eq:Js}
J_s^\gtrless(z,E) = m_s\int_{0}^{E-U_s(z)}\rmd TF_s^\gtrless(z,E,T)~,
\end{equation}
in terms of which the macroscopic fluxes required for the characteristic become
\begin{equation}
\label{eq:js}
j_s(z) = \int_{U_s(z)}^\infty\frac{ \mathrm{d} E}{8\pi^2}\left[J_s^>(z,E)- J_s^<(z,E)\right]~.
\end{equation}
	
\subsection{Numerical strategy}\label{sec:Numerics}
In the following, we give a sketch of the numerical strategy used for solving the kinetic problem 
stated in the previous subsection. Due to the changes in the boundary and matching conditions,
the strategy differs somewhat from the one used previously~\cite{RBF20}. In particular, the procedure for 
establishing selfconsistency between the plasma and the solid side of the interface is different due to 
the reflectivity of the interface. The isotropy of the collision integrals enables us moreover to discretize 
a much larger energy domain.

Besides the distribution functions $F_s^\gtrless(z,E,T)$ and the electric potential profile $U(z)$, 
three energy parameters $U_W$, $U_I$, and $U_{SC}$ and two density parameters 
$n^\textrm{LM}_e$ and $n^\textrm{LM}_i$ have to be selfconsistently determined for prescribed 
plasma density $n_p$ and external voltage $U_B$. The five equations required for it are,
the charge neutrality condition at $z=z_p$, providing two equations, $n_p=n_e(z_p)=n_i(z_p)$, the 
electric matching condition~\eqref{eq:matching}, the generalized Bohm criterion~\eqref{eq:GBC}, and 
the condition~\eqref{VoltageDrop} following from the definitions of the potential drops at the interface. 
The net flux through the interface,
\begin{align}
j\equiv j_*(0)-j_h(0)=j_e(0)-j_i(0)~, 
\end{align}
due to flux conservation identical to the flux anywhere in the device, is then obtained 
as a function of $U_B$. It will be however numerically advantageous to specify 
$U_{SC}$ instead of $U_B$ and to initially determine $j(U_{SC})$ from which 
$j(U_B)$ follows straight by applying Eq.~\eqref{VoltageDrop}.
	
For the numerical implementation of the selfconsistent calculation of the current-voltage characteristic
we rewrite the Boltzmann equations~\eqref{eq:BEQ} for the charge carriers inside the semiconductor  
in integral form (suppressing the parametrical dependencies on $E$ and $T$)~\cite{RBF20}, 
\begin{subequations}\label{eq:sol_it}
\begin{multline}\label{eq:sol>it}
F_s^{>}(z) = \xi_s(z,z-\Delta) F_s^>(z-\Delta)\\
 + \int_{z-\Delta}^{z} \mathrm{d} z' \frac{\Phi_s(z')}{v_s(z')} \xi_s(z,z')
\end{multline}
and
\begin{multline}\label{eq:sol<it}
F_s^{<}(z) = \xi_s(z +\Delta,  z) F_s^<(z + \Delta) \\
+ \int_{z}^{z + \Delta}   
\mathrm{d} z'\frac{\Phi_s(z')}{v_s(z')} \xi_s(z',z)
\end{multline}
\end{subequations}
\noindent with the integrating factor 
\begin{equation}\label{eq:I}
\xi_s(z,z') = \exp\left( -\int_{z'}^{z} \mathrm{d} \bar z\frac{\gamma_s(\bar z)}{v_s(\bar z)} \right)~
\end{equation}
and $s=*,h$.

To avoid the integrable divergences of $1/v_s(z,E,T)$ at $T=E-U_s(z)$, that is, at the turning points
for the perpendicular motion, where the charge carriers move parallel to the interface, it is 
convenient to perform a coordinate transformation from $z$ to \mbox{$X_s = \sqrt{E-U_s(z)-T}$}. Using 
\mbox{$\partial U_h/\partial z =\mathcal{E}$},\mbox{ $\partial U_*/\partial z =-\mathcal{E}$}, and 
$v_s = 2X_s/\sqrt{m_s}$, we find
\begin{subequations}
\label{eq:substitution}
\begin{align}
\frac{\rmd z}{v_h} &= -\sqrt{m_h} \frac{\rmd X_h}{\mathcal{E}}~,\\
\frac{\rmd z}{v_*} &= \sqrt{m_*} \frac{\rmd X_*}{\mathcal{E}}~,
\end{align}
\end{subequations}
\noindent which enables us to rewrite the $z-$integrals as $X_s-$integrals. If the semiconducting layer 
is not too thick, the electric field $\mathcal{E}$ is finite in the whole integration domain. Thus, 
using this substitution we can avoid diverging integrands in the numerical solution of the Boltzmann 
equations. 
	
After the transformation, the coordinates $E$, $T$, and $X_s$ are discretized, with the discretization 
kept fixed during the iteration. Through the domain of $X_{*,h}$, we thus prescribe the value $U_{SC}$, 
although it is actually a parameter to be determined selfconsistently for given $U_B$. Instead of 
$U_{SC}$, we use $U_B$ as a derived parameter, which allows a better handling of the turning points, 
where $v_{*,h}=0$ and $F_{*,h}^>=F_{*,h}^<$. Due to Eq.~\eqref{VoltageDrop} this is permissible.

The remaining four parameters, $U_I$, $U_W$, $n_e^\mathrm{LM}$, and $n_i^\mathrm{LM}$ are determined
from the charge neutrality at $z=z_p$, bringing in two equations and yielding $n_{e,i}^\mathrm{LM}$ 
in terms of $U_W$ and $U_I$, the matching condition~\eqref{eq:matching}, and the generalized Bohm 
criterion~\eqref{eq:GBC}. The latter two provide at the end two coupled equations for $U_I$ and $U_W$ 
which have to be solved selfconsistently with the Boltzmann equations and the Poisson equation on
both sides of the interface.

To get the two equations, we relate the potential drop $U_{SC}$ to $\mathcal{E}(z)$ by the 
integral
\begin{equation}\label{eq:U_SC}
U_{SC} = \int_{z_0}^{0}\rmd z \mathcal{E}(z)~.
\end{equation}
Inserting $\mathcal{E}(z)$ from Eq.~\eqref{eq:U'SC}, setting $n_A=n_D=0$, since we consider in the next 
section an undoped germanium layer, and solving for $\mathcal{E}(0^-)$ yields
\begin{equation}
\label{E_left}
\mathcal{E}(0^-) = \left[ U_{SC} + 
\frac{8\pi}{\varepsilon}\int_{z_0}^{0}\rmd z\int_{z}^{0}\rmd z'(n_*(z')-n_h(z'))\right]/z_0~,
\end{equation}
that is, the electric field at $z=0$ necessary to produce, with the net charge distribution $n_*(z)-n_h(z)$, 
the potential drop $U_{SC}$ over the width of the semiconductor. 

The electric field $\mathcal{E}(0^+)$, on the other side of the interface, in turn can be obtained 
from Eq.~\eqref{eq:U'P}. Inserting the electron and ion densities arising from the electron and ion distribution 
functions $F_{e,i}^\gtrless$, which can be largely worked out analytically~\cite{RBF20}, 
we get   
\begin{align}
\label{E_right}
\mathcal{E}(0^+)=\sqrt{16\pi\left(G_e + G_e^> + G_i\right)}
\end{align}
with 
\begin{widetext}
\begin{align}
G_e   &= \frac{n_p-n_e^>}{1 + \mathrm{erf}\left(\sqrt{-\tilde{\beta}_e \tilde{U}_I}\right)}
      \left[\exp\left( \tilde{\beta}_e \tilde{U}_I \right)\left(1 + 2\sqrt{-\frac{\tilde{\beta}_e \tilde{U}_I}{\pi}}\right) 
      - 1 - \mathrm{erf}\left(\sqrt{-\tilde{\beta}_e \tilde{U}_I}\right)  \right]~,\\
G_e^> &= \frac{m_e^{3/2}}{32\pi^2}
      \int_{0}^\infty \rmd E\int_{0}^{E} \rmd T F_e^>(z_p,E,T) 
      \left( \sqrt{E+U_I -T}-\sqrt{E-T} \right)~, \\
G_i   &= \frac{n_p}{\mathrm{erfc}\left(\sqrt{\tilde{U}_W}\right)}
      \bigg\{\mathrm{erfc}\left(\sqrt{\tilde{U}_W}\right) - \exp\left(-\tilde{U}_I\right) 
      \mathrm{erfc}\left(\sqrt{\tilde{U}_I-\tilde{U}_W}\right)
      + 2 \exp\left(-\tilde{U}_W\right)\left( \sqrt{\frac{\tilde{U}_W}{\pi}} 
      - \sqrt{\frac{\tilde{U}_W-\tilde{U}_I}{\pi}} \right) \bigg\}~,
\end{align}
\end{widetext}
where we used $U_e(z_p)=0$, because of the choice of the reference point for the potentials,
set $\tilde{U}_{I,W} = U_{I,W}/k_B T_i$, and $\tilde{\beta}_e = k_B T_i/k_B T_e$. The symbols 
$\mathrm{erf}(x)$ and $\mathrm{erfc}(x)=1-\erf(x)$ denote the error and complementary error 
function.

Multiplying Eq.~\eqref{E_left} by the dielectric constant $\varepsilon$ and equating it with
Eq.~\eqref{E_right}, an equation is obtained relating $U_{SC}$ to $U_W$, $n_p$, and $U_I$. Since 
$n_p$ and $U_{SC}$ are external parameters, we thus have a relation between $U_W$ and $U_I$, as 
required. For a perfectly absorbing interface $G_e^>=0$. 
	
A second relation between $U_W$ and $U_I$ follows from the generalized Bohm criterion. 
Expressing again $n_{e}$ and $n_i$ in terms of the distribution functions $F_{e,i}^\gtrless$, 
and using the definitions introduced above, we obtain
\begin{align}\label{eq:GBC_f}
\frac{f_+\left(-\tilde{\beta}_e\tilde{U}_I\right)}{k_B T_e}\left(1 - \frac{n_e^>}{n_p}\right) 
+ \frac{\Delta_e^>}{n_p} - \frac{f_-\left(\tilde{U}_W\right)}{k_B T_i} = 0 
\end{align}
with
\begin{align}
f_\pm (x) &= \frac{e^{-x}}{\sqrt{\pi x}(1 \pm \mathrm{erf}(\sqrt x))} \pm 1~
\end{align}
and 
\begin{align}
\Delta_e^> &= -\frac{m_e^{3/2}}{32\pi^2}\int_{0}^\infty\rmd E\int_0^{E} \rmd T \frac{F_e^>(z_p,E,T)}{(E-T)^{3/2}}~,\\
n_e^> &= \frac{m_e}{8\pi^2}\int_{0}^\infty\rmd E\int_0^{E} \rmd T \frac{F_e^>(z_p,E,T)}{\sqrt{E-T}}~,
\end{align}
where $U(z_p)=0$ has been used once again. We thus have a second
equation connecting $U_W$ and $U_I$. Note, through Eq.~\eqref{eq:BC_e}, the electron distribution function 
$F_e^>$ on the plasma side depends on the electron distribution function $F_*^>$ inside the solid, providing 
an additional feedback of the solid to the plasma, in addition to the electric matching~\eqref{eq:matching}.
For a perfectly absorbing interface this kind of feedback is absent.

With the parameters $U_W$ and $U_I$, the particle fluxes and hence the source 
functions $S_{*,h}^>$ are fixed. Inserted in the boundary conditions~\eqref{eq:BC_h}  
and~\eqref{eq:BC_*}, the distribution functions for valence band holes and conduction band electrons 
are obtained from Eqs.~\eqref{eq:sol_it}, with the substitutions~\eqref{eq:substitution}, from which 
follow also the densities $n_{*,h}(z)$ according to Eq.~\eqref{eq:n}, which in turn can be fed into the 
Poisson equation~\eqref{eq:U'SC} to yield a new potential $U(z)$ to be used in the next iteration
step. For semiconductors with high intrinsic densities, such as germanium, it turns out that the 
change in densities caused by the injected charge carriers is negligible compared to the Maxwellian 
background of intrinsic carriers. It is thus possible to neglect it in the source term of the Poisson 
equation, reducing thereby the calculational costs substantially. The densities plotted in 
Fig.~\ref{fig:macroscopic_plots} verify a posteriori the validity of this simplification.

The iteration starts with constant Maxwellian distributions in accordance to the boundary 
conditions~\eqref{eq:BCallfs}, but for conduction band electrons continued to energies which are not 
accessible at $z=z_0$ due to the higher value of $U_*$. Successively, the distribution functions 
are updated by Eq.~\eqref{eq:sol_it}, as are the parameters $U_W$ and $U_I$ by Eq.~\eqref{eq:matching},
expressed in terms of Eqs.~\eqref{E_left} and~\eqref{E_right}, and Eq.~\eqref{eq:GBC_f}. In contrast 
to the perfectly absorbing interface~\cite{RBF20}, the plasma parameters 
now change slightly in each iteration step, which in turn changes also the matching conditions at 
the interface. The updating is repeated until convergence is reached. It should be noted that 
the conduction band electron and valence band hole distribution functions also enter
the in-scattering collision integrals $\Phi_s$ in Eqs.~\eqref{eq:sol_it}. Hence, even without 
the coupling to the plasma and changing boundary conditions, an iteration is required to solve
the Boltzmann equations inside the germanium layer. This is also the case when the 
Grinberg-Luryi approach~\cite{GL92} is applied to semiconductor device modeling~\cite{DP98,KH02}. 

\begin{figure}[t]
\includegraphics[width=\linewidth]{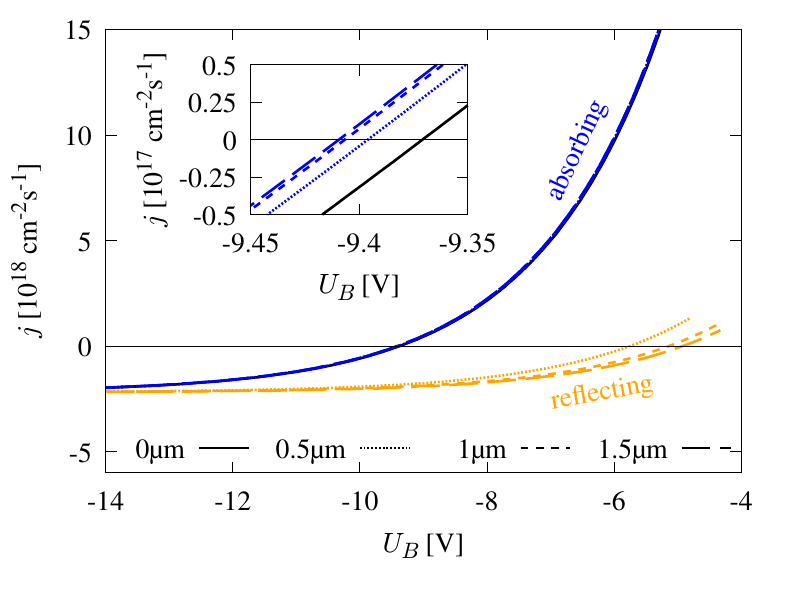}
\caption{(color online) Current-voltage characteristic across a device consisting of a germanium
layer sandwiched between an argon plasma and an Ohmic contact. The orange lines belong to the reflecting
plasma-germanium interface whereas the blue lines show data for an interface which perfectly absorbs 
electrons from the plasma and lets them never return to it. Different layer thicknesses are considered. For 
comparison, we also plot the characteristic of a perfect absorber without a germanium layer, that is, 
for $U_{SC}=0$ (solid black line, hardly seen in the main panel). The inset shows for the perfectly 
absorbing interfaces the zero-crossings of the net flux. The germanium layer shifts the crossings to 
lower voltages because of the potential drops $U_{SC}$.}
\label{fig:UI}
\end{figure}

\section{Results}\label{sec:results}

This section discusses the numerical results obtained for an argon plasma in contact with a germanium
layer. The material parameters are given in Table~\ref{tab:parameters}. We split the discussion into 
three parts, depending on the perspective from which the device shown in Fig.~\ref{fig:concept} is 
analyzed. First, regarding it as part of an electric circuit, we present the current-voltage 
characteristic in subsection~\ref{sec:BP}. Then, we proceed to discuss in subsection~\ref{sec:MacP}
the spatial profiles of the electric potential and the species' densities as well as fluxes. 
Finally, in subsection~\ref{sec:MicP}, we turn to the energetically and spatially resolved 
distribution functions of the charge carriers inside the semiconducting layer.
	
\subsection{Electric picture}\label{sec:BP}

\begin{figure}[t]
\includegraphics[width=\linewidth]{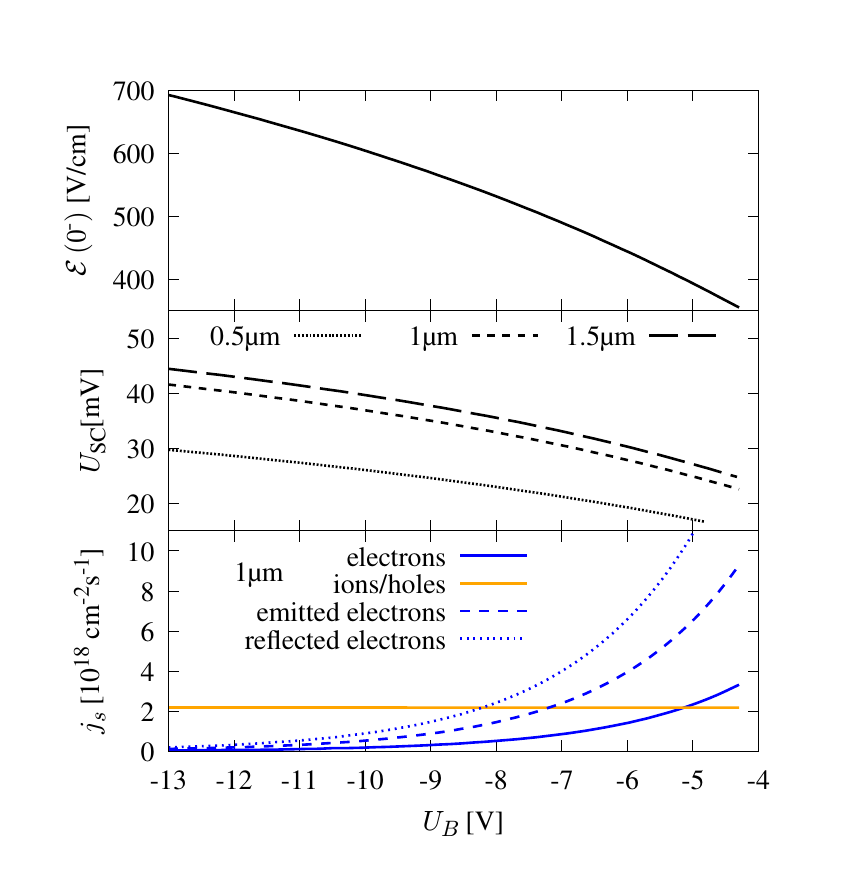}
\caption{(color online) As a function of $U_B$ and for the reflecting plasma-solid interface, 
the electric field at the interface (top panel), 
the potential drop $U_{SC}$ across the germanium layer (center panel), and the electron and 
ion/hole fluxes flowing from the plasma towards the Ohmic contact (bottom panel). Reflected 
and emitted electrons refer, respectively, to electrons quantum-mechanically reflected at 
the potential step connected with the plasma-solid interface and electrons emitted back to 
the plasma due to collisions inside the semiconductor. While the electric field is identical 
for the layer thicknesses considered, $U_{SC}$ changes with the thickness. Electric field 
and $U_{SC}$ are effectively identical to what one obtains for absorbing matching conditions. 
Fluxes are given only for the case of a $1\,\mu$m thick germanium layer.}
\label{fig:U-curves}
\end{figure}

A biased plasma-solid interface can be considered as part of an electric circuit and thus as an 
electric device characterized by a current-voltage characteristic, that is, the net flux  
(current density) flowing through the system as a function of the bias voltage $U_B$. For 
germanium layers of different thicknesses in contact with an argon plasma, the characteristics are 
shown in Fig.~\ref{fig:UI}. 

The germanium layers are terminated by an Ohmic contact, which is not 
further characterized in the model. It merely acts as a sink for any flux carrying particles 
reaching the semiconductor-metal interface. For $U_{SC}=0$, that is, without the semiconducting layer,
the system is thus a perfectly electron absorbing Langmuir probe~\cite{ML26, Cherrington82,Lam65}. 
Current-voltage characteristics are given for different thicknesses of the germanium layer, for both 
the electron reflecting and the perfectly absorbing semiconductor-plasma interface. In the latter, the 
matching conditions~\eqref{eq:BC_*} and~\eqref{eq:BC_e} are replaced by the assumption that every 
electron impinging on the interface from the plasma is absorbed by the germanium layer and every 
electron reaching the interface from the inside of the layer is specularly reflected. The 
boundary conditions then take a form similar to Eqs.~\eqref{eq:BC_i} and~\eqref{eq:BC_h}, with 
the injection term~\eqref{eq:S_*} normalized such that the electron flux is conserved across the 
interface. 
	
Within the perfect absorber model, there are hardly any differences in the current-voltage 
characteristics of the Ohmic and the semiconducting plasma-solid interface (black and blue 
lines in Fig.~\ref{fig:UI}). The characteristics of the latter are 
only shifted to lower voltages because of the voltage drops $U_{SC}$ inside the semiconductor, as 
can be seen in the inset of the figure for different film thicknesses. The shifts are a few tens 
of mV, with a thicker layer giving rise to a larger shift. If the layers were thick enough to host the 
whole negative space charge, the shifts would saturate, because of the vanishing electric field deep
inside the semiconductor. For the thicknesses shown in Figs.~\ref{fig:UI} this is however not yet the 
case. Since, as discussed in the next paragraph, the voltage drops $U_{SC}$ across reflecting and 
absorbing germanium layers of the same thickness turn out to be essentially identical, because $n_e^>$ 
and $\Delta_e^>$ are rather small, and hence effectively negligible in Eq.~\eqref{eq:GBC_f}, the large 
shift between the blue and orange lines visible in the main panel is a consequence of the 
reflected and emitted electron fluxes, which can be rather significant. From a broader perspective, 
the results shown in Fig.~\ref{fig:UI} imply that modifications of the surface of a Langmuir probe, 
for instance, by an oxide film, affects the current-voltage characteristic only by a small amount if 
the perfect absorber assumption holds. However, if this is not the case, the characteristic depends 
on the film. In particular, the zero-crossing, that is, the floating potential, would depend strongly 
on the emissive properties of the film.

To gain more insights, we plot in Fig.~\ref{fig:U-curves}, as a function of the bias voltage $U_B$, the 
electric field at the interface, the potential drop $U_{SC}$ across the semiconducting layer, and the 
electron and hole fluxes. The electric field at the interface is independent of the thickness of the 
layers, only the potential drop depends on it. The fluxes are shown representatively only for a 
$1\,\mu\mathrm{m}$ thick semiconducting layer, with electron fluxes split into the contributions arising 
from electrons reflected at the potential step and electrons emitted from the semiconductor, that is, 
electrons which made it into the layer but are backscattered by electron-phonon collisions inside it. The 
two contributions arise, respectively, from the first and second term of Eq.~\eqref{eq:BC_e}. 

While the reflected flux is independent of the thickness of the layer, since it only depends on the reflection 
coefficient $R(E,T)$, the emitted flux increases with the thickness of the layer, leading to smaller total 
net fluxes towards the Ohmic contact, as can be also seen in Fig.~\ref{fig:UI}. This may be an artifact of 
the simple model we employ for the germanium layer. Flux-carrying electrons, having typically large energy,
as we will see in subsection~\ref{sec:MicP}, undergo in this model only a few collisions while passing through 
the layer. The thicker the layer, the higher is thus the chance of an electron belonging to this group for 
being backscattered. Hence, the emitted flux increases with layer thickness.  

The electron fluxes decay exponentially the more negative the bias voltage is, since the fraction 
of the Maxwellian distribution of plasma electrons contributing to the flux by overcoming the sheath 
potential $U_I$ decreases. Only few electrons in its high energy tail contribute to the flux across 
the interface and have thus a chance to get reflected or emitted. The emitted and reflected 
electrons, encoded in $\Delta_e^>$ and $n_e^>$, modify thus~\eqref{eq:GBC_f} only weakly and are 
hence negligible, as it is also the case in the source term of the Poisson equation, where emitted
and reflected electrons are dominated by the plasma electrons belonging to the low energy part of 
the Maxwellian.

\subsection{Macroscopic picture}\label{sec:MacP}

Besides regarding the interface as an electric device, characterized by a current-voltage characteristic,
it is also instructive to consider it as a system of charged particles, characterized by density, flux, and 
potential profiles. We limit the discussion in this and the next subsection to the reflecting interface 
at the floating point, where the electron and ion/hole fluxes are equal, noting, however, that 
for different bias voltages the fundamental observations are similar, differing only by the numerical values.

\begin{figure}[t]
\includegraphics[width=\linewidth]{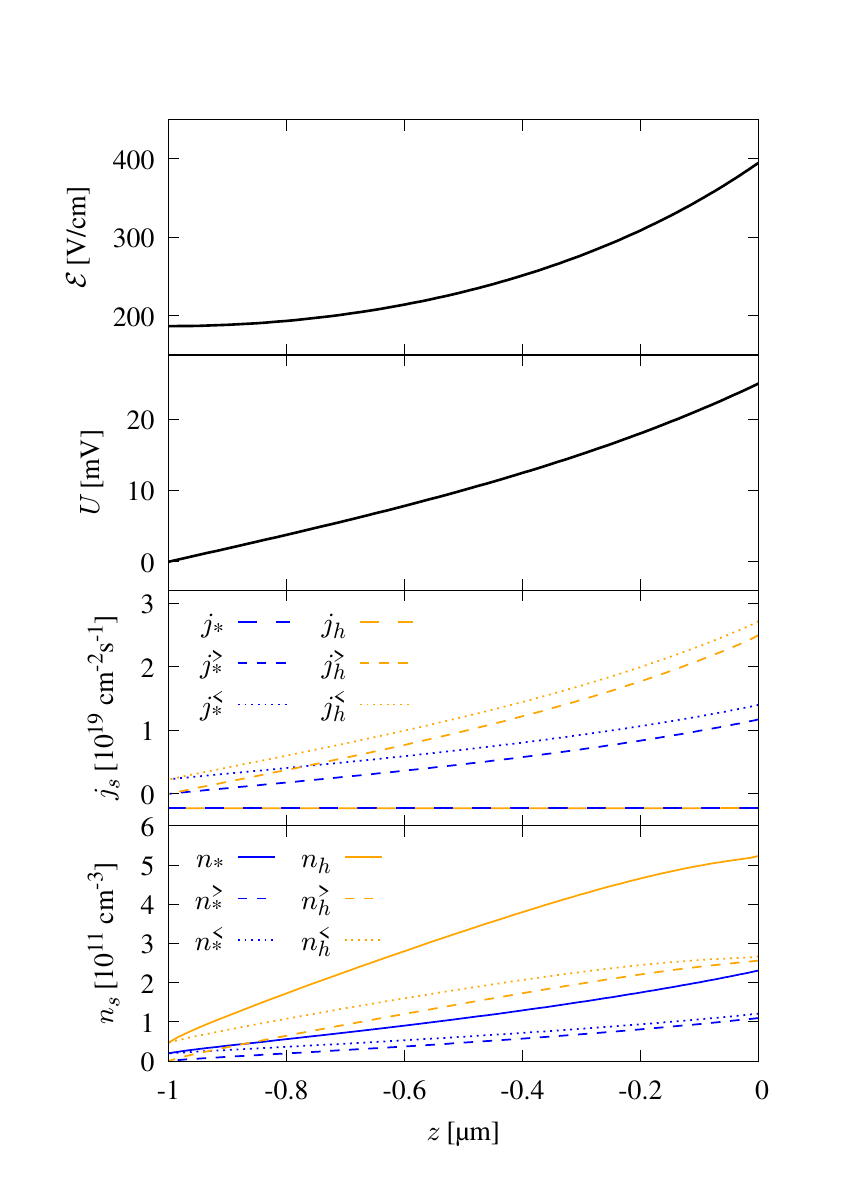}
\caption{(color online) Macroscopic properties for the reflecting $1\,\mu\mbox{m}$ thick interface
at the floating point, for which $U_I=5.1\,$eV, $U_{SC}=0.025\,$eV, and $U_W=0.97\,$eV.
Top to bottom: Electric field, electric potential (shifted so that 
$U(z_0)=0$), fluxes, and densities. The last two are only given for the injected surplus electrons 
and holes, the Maxwellian contributions due to the intrinsic carriers, giving rise to the double layer 
shown in the inset of the top panel, are subtracted. In addition to the net fluxes and densities per
species, we also plot the direction-resolved quantities.}  
\label{fig:macroscopic_plots}
\end{figure}

Figure~\ref{fig:macroscopic_plots} shows from top to bottom the spatial profiles for the electric field, 
the electric potential, the fluxes, and the densities of the injected carriers for an interface with a 
$1\,\mu\mathrm{m}$ thick germanium layer. As it was the case on the plasma side, the injected carriers, 
for which we show the fluxes and densities, are also negligible in the source term of the Poisson equation 
on the solid side of the interface. The solid-bound parts of the electric field and potential shown in 
the top two panels are thus determined by the intrinsic charge carriers. Hence, the solid-bound part of the 
electric double layer plotted in the inset of the top panel, is the result of the selfconsistent distribution 
of the intrinsic charge carriers in the electric field arising due to the interface. For germanium, the 
intrinsic density at room temperature is $2\cdot 10^{13}\,\textrm{cm}^{-3}$, and thus much larger than the 
net density of the injected carriers, which, according to the bottom panel of Fig.~\ref{fig:macroscopic_plots}, 
is even at $z=0$ only only around $3\cdot 10^{11}\,\textrm{cm}^{-3}$, and hence about two orders of magnitude 
smaller. The simplification we made for the solution of the Poisson equation is thus valid. It not only 
stabilizes the iteration process, but allows also for an efficient implementation, which reduces the 
computation time by about two orders of magnitude. 
	
The electric potential, shown in the second panel from the top of Fig.~\ref{fig:macroscopic_plots}, is 
shifted such that $U(z_0)=0$. The potential drop across the germanium layer $U_{SC}=25\,$meV can thus be 
read off directly at $z=0$. It can be also immediately seen that the electric field 
is the derivative of the electric potential. The fluxes, finally, are shown in the third panel from the 
top. In addition to the net fluxes, $j_*$ and $j_h$, which are conserved and hence independent of $z$, 
left- and right-moving fluxes are also plotted. They decay exponentially with decreasing $z$. The net 
flux is the difference of the right- and left-moving fluxes. It is negative, showing that more 
electrons and holes are moving towards the Ohmic contact than in the other direction. The conservation 
of the net flux is a consequence of the modeling of the germanium layer, which contains only 
particle-conserving collisions with phonons. Had we incorporated also particle non-conserving scattering 
processes, such as radiative and non-radiative electron-hole recombination or impact ionization, the net 
fluxes of each species would not be conserved. However, to have flux conservation as an important indicator 
of the numerical accuracy of the implementation of the selfconsistency scheme (it is not explicitly kept 
constant in the iteration scheme), we neglected in this exploratory work particle non-conserving scattering 
processes. 
	
\subsection{Microscopic picture}\label{sec:MicP}

We now turn to the microscopic picture as it arises from the distribution functions satisfying the 
two coupled sets of Boltzmann-Poisson equations. The electric and macroscopic pictures contain
this information only in an integral manner. Now, we take an energy-resolved look at the 
kinetics of the charge carriers across the interface. The dependence of the distribution functions 
on the total energy $E$ and the lateral kinetic energy $T$ enables us to visualize energy relaxation 
due to collisions with phonons as well as the motion of the charge carriers parallel to the interface.

\begin{figure}[t]
\includegraphics[width=\linewidth]{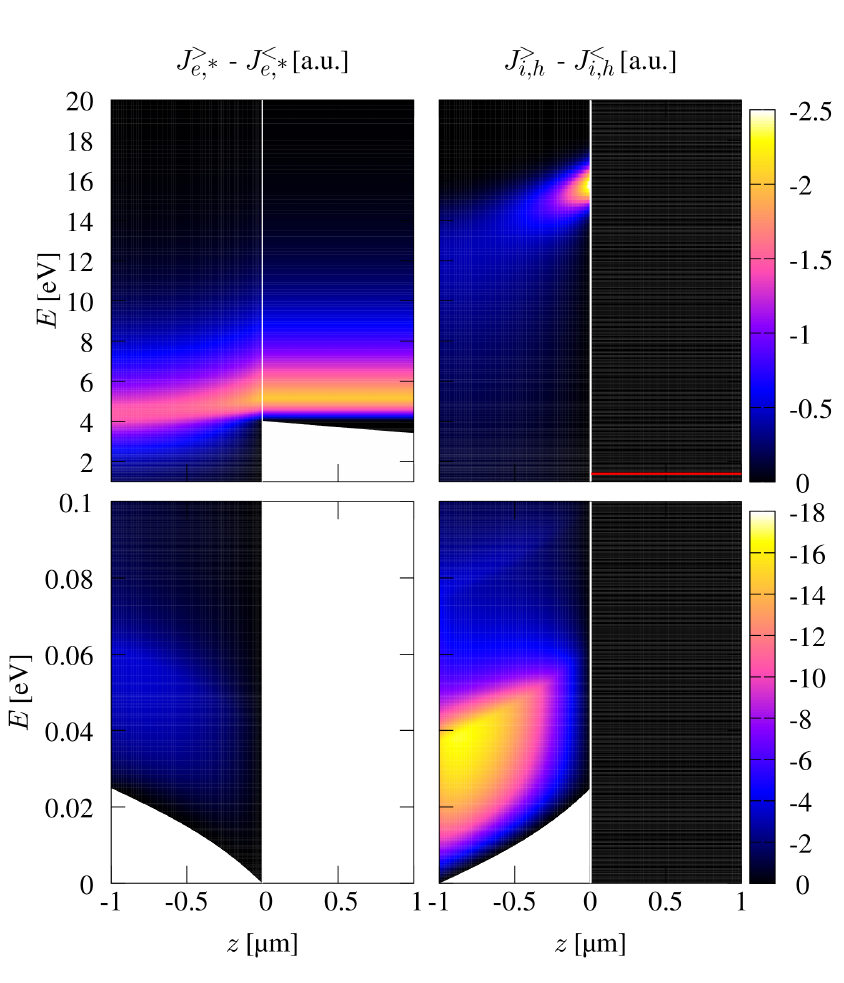}
\caption{(color online) In arbitrary units, the energy-resolved fluxes across 
the interface with a $1\,\mu\mathrm{m}$ thick germanium layer. The interface is at the 
floating point with potential drops given in the caption of Fig.~\ref{fig:macroscopic_plots}. 
On the left (right), the net electron and hole (electron and ion) fluxes inside 
the germanium layer (plasma) are shown, with high and low energy contributions plotted in separate 
panels. The ion flux, located at $U_I+U_W+U_{SC}-\chi-E_g=1.39\,$eV, is shown by the red line (not 
belonging to the color scale). Its energetic spread is negligible on the scale of the 
other fluxes. The energy scale is chosen such that within the germanium layer $E=0$ occurs at the 
minimum of $U_*$ and $U_h$, respectively.} 
\label{fig:flux_3d}
\end{figure}

Let us first look at the energy-resolved net fluxes per species defined in Eq.~\eqref{eq:Js}. For 
an interface with a $1\,\mu\mathrm{m}$ thick germanium layer at the floating point, the fluxes 
flowing from the plasma towards the Ohmic contact are plotted in Fig.~\ref{fig:flux_3d}. On the 
left (right), fluxes inside the germanium layer (plasma) are shown. While the species' energy-integrated 
net fluxes are constant, the energy-resolved net fluxes display a rich behavior along the germanium layer.

The net electron flux approaching the interface from the plasma side splits inside the solid into
a high- and a low-energy component. The former comprises electrons remaining close to the energy 
where they have been initially injected. Because the germanium layer is rather thin, a large 
number of electrons reaching the Ohmic contact suffer only a few electron-phonon collisions. 
Hence, they loose only a small amount of energy and remain energetically high in the conduction 
band. Yet, some electrons scatter often enough to end up at the band minimum. There is thus also 
a net flux at low energy. In absolute numbers, this is even larger than the one at high energies
(note the different scales for high and low energies). However, its contribution to the macroscopic 
net flux is rather small because the energy range over which it is integrated is rather narrow. 
The main contribution to the net total electron flux arises thus from high-energy electrons.
In numbers, we find $98\%$ of the total electron flux to be carried by the hot electrons, with 
energies above $0.1\,\mathrm{eV}$, and only $2\%$ by the low energy electrons.

The net flux of valence band holes, arising from the neutralization of ions, hitting the interface 
with thermal kinetic energy, shows qualitatively the same behavior. Holes, injected high up in the 
valence band at $E=I_{\rm inj}$, relax due to collisions with phonons to the bottom of the band.  
Quantitatively, however, the situation is different. Due to the larger effective mass, 
holes loose energy due to collisions with phonons more efficiently than electrons. The scattering 
rate $\gamma_{s}$, for instance, defined in Eq.~\eqref{eq:gamma}, is proportional to $m_s^{3/2}$. Hence,
it increases with the effective mass. The same holds for the in-scattering parts of the collision
integral $\Phi_{s}$. Thus, the larger the effective mass of the charge carriers, the smaller is
the inelastic mean free path, and hence the spatial scale required to loose a substantial amount
of energy. The $1\,\mu\mathrm{m}$ thick germanium layer provides apparently enough space for a 
significant number of holes to relax to the bottom of the band and to give rise to a rather 
pronounced hole flux at low-energies. Integrated over energy, the low-energy flux 
provides about $20\%$ to the total net flux. The high-energy flux is hence still dominant but 
not as dominant as in the case of electrons. 

Energy relaxation of electrons and holes stops when their energies, measured from the bottom of 
the conduction and valence band, respectively, are less then the phonon energy. There is thus a 
threshold for energy relaxation, leading to a kind of energy backlog in the fluxes up to a phonon 
energy above the band minima. By absorbing phonons, carriers caught in the backlog can reach higher 
energies. For holes, this is clearly visible in Fig.~\ref{fig:macroscopic_plots}. It leads to the 
feature in the bottom right panel around $E=0.08\,\mathrm{eV}$, which is 
roughly twice the phonon energy. After the injected holes emitted several hundred phonons, each 
carrying away $\hbar\omega_0=0.037\,$eV, their energies fall below the threshold. Collisions are 
then less frequent, as the emission (second) term in the scattering rate~\eqref{eq:gamma} disappears. 
The relatively high population of the holes just below the phonon energy makes however the absorption 
process operative in the collision integrals leading to the feature in the hole flux around 
twice the phonon energy. 

\begin{figure*}[t]
\includegraphics{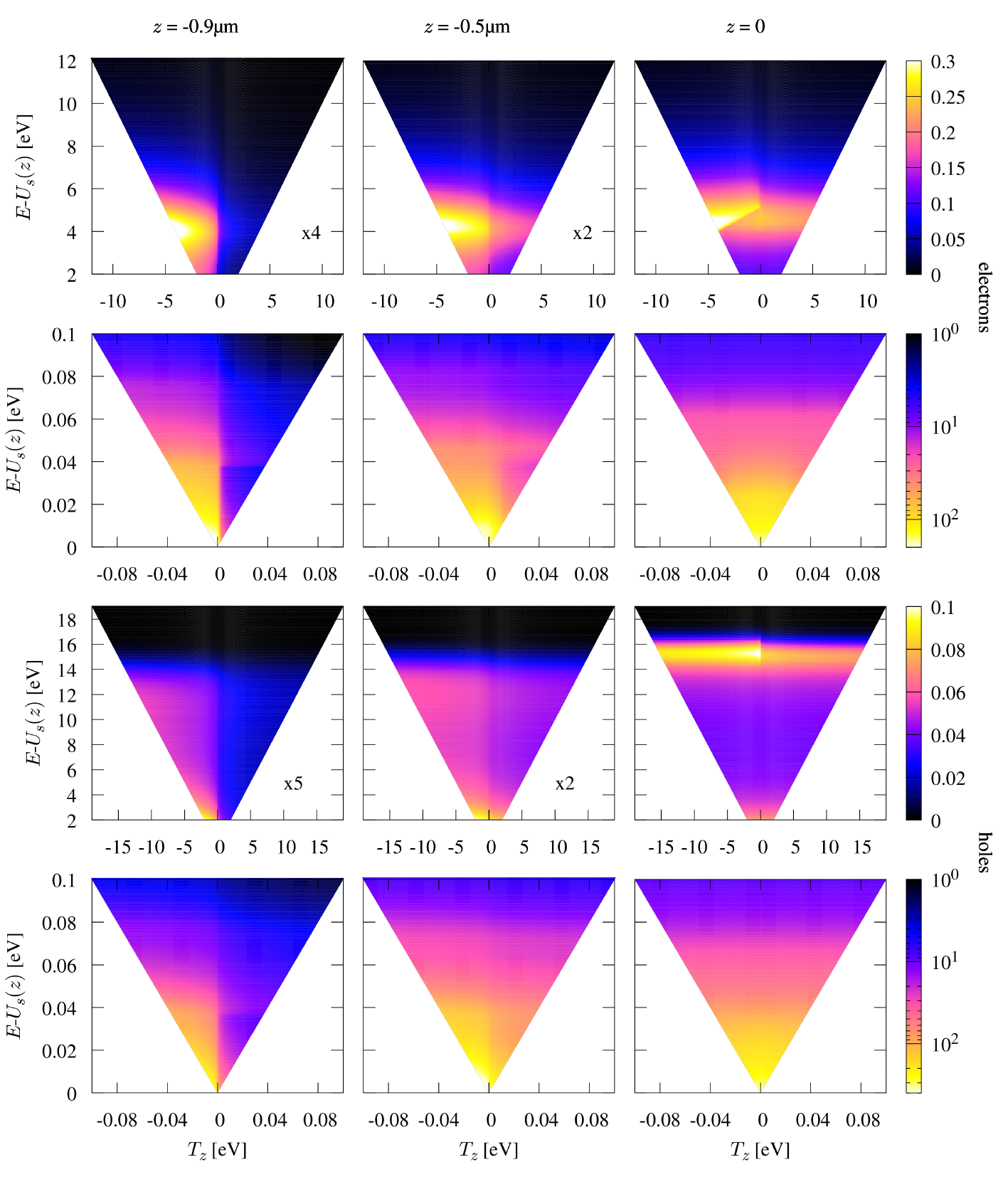}
\caption{(color online) Distribution functions in arbitrary units for electrons (top two rows) and holes (bottom 
two rows) at different spatial locations of a $1\,\mu\mathrm{m}$ thick nearly floating germanium layer. Regions 
of high and low energy are shown in separate panels. On the horizontal axis, positive values of $|T_z|=E-U_s-T$ 
show $F^>_s$ and negative values depict $F_s^<$. The triangular shape is due to the energy restriction $0<T<E-U_s$. 
Note the logarithmic (linear) scale for low (high) energies. At $z=-0.9\,\mu$m and $z=-0.5\,\mu$m, the high energy 
distributions are multiplied by the factors given in the panels to utilize the same scale. 
The potential drops are as in Figs.~\ref{fig:macroscopic_plots} and~\ref{fig:flux_3d}.}
\label{fig:distributions}
\end{figure*}

Finally, we discuss for the $1\,\mu\mathrm{m}$ thick germanium layer the distribution functions
$F_{*,h}^\gtrless(z,E,T)$. In addition to the effects already present in the energy-resolved net fluxes,
there are now also features due to the lateral motion of the charge carriers. The distribution functions
are plotted in Fig.~\ref{fig:distributions}, with the two top (bottom) rows showing the data for conduction 
band electrons (valence band holes). Three spatial locations are considered and data for high and
low energies are plotted in separate panels. The lateral coordinate of the plots encodes by its sign 
also the direction of motion, and is chosen such, that for $|T_z|=E-U_s-T=0$ it describes the turning 
points. Note the linear scale for high and the logarithmic scale for low energies. As for the macroscopic 
densities plotted in Fig.~\ref{fig:macroscopic_plots}, the Maxwellian background of intrinsic carriers 
is not included in the data. Only the distribution functions for the surplus electrons and holes arising 
from the plasma are shown.

In the high energy parts of the distribution functions right at the interface, at $z=0$ (right column), 
the source functions $S_*^<$ and $S_h^<$, describing the injection of electrons and holes, can be clearly 
identified. Due to the collisions with phonons there is also a noticeable portion of right-moving 
charge carriers, as can be seen from the rather large numerical values for $F_{*,h}^>$. For electrons, 
this group of charge carriers eventually leads to the flux of emitted electrons, if they also make it 
through the potential barrier, and hence to secondary electron emission. As a consequence of energy 
relaxation, Maxwellian distributions are established for both species at low energies. Indeed, 
the low-energy parts of the distribution functions are rather homogeneous in the lateral direction 
and exponentially decaying with total energy, as it should be for Maxwellians. As a consequence 
of the Maxwellian distribution at low energies, there is no low-energy flux present at $z=0$, 
in accordance with the data shown in Fig.~\ref{fig:flux_3d}.

In the center of the germanium layer, at $z=-0.5\,\mu$m (center column), the sharp features of the 
source functions are softened due to collisions with phonons. A surplus motion to the left is also
building up, most distinctly for conduction band electrons, and weaker for valence band holes. At
low energies, the surplus of left-moving charge carriers remains, but it is now somewhat more dominant 
for holes. Due to the imbalances, the distribution functions at low energy start to deviate from
Maxwellian distributions. As a result, low-energy fluxes are building up, as can be also seen in
Fig.~\ref{fig:flux_3d}. The behavior of the energy-resolved net fluxes can thus be explained by 
the changes in the distribution functions caused by the scattering processes.

The distribution functions at \mbox{$z=-0.9\,\mu$m} (left column), finally, are already dominated by the boundary
condition $F_{*,h}^>=0$, set for the injected carriers at the interface to the Ohmic contact. For both species,
the motion is heavily biased to the left, irrespective of the energy. While right-moving electrons and holes 
are strongly suppressed, there is a faint feature in $F_{*}^>$ and  $F_{h}^>$ slightly above the phonon 
energy which arises from collisions with phonons. The features of the source functions are still visible
in the high-energy parts of $F_*^<$ and  $F_h^<$, albeit severely washed out, and at low energies 
$F_*^<$ and  $F_h^<$ remain Maxwellian. It should be noted that at high energies the absolute values of 
the distribution functions decay with $z$ approaching $z_0$. The ratio, however, of left to right moving 
distributions grows. Hence, the net flux at high energies stays high, even though the individual distribution
functions $F_{*,h}^\gtrless$ for left- and right-moving carriers decay with decreasing $z$, just like the 
direction-resolved fluxes $j_{*,h}^\gtrless$ shown in Fig.~\ref{fig:macroscopic_plots}.

\section{Conclusion}\label{sec:conclusion}

We presented a kinetic description of ambipolar charge transport across a biased plasma-solid 
interface consisting of a semiconducting germanium layer sandwiched between an argon plasma and an 
Ohmic contact. The electron-hole plasma within the semiconductor is coupled to the electron-ion plasma 
in front of it through matching conditions for the distribution functions and the electric field 
at the interface. Argon ions impinging on the germanium layer create holes in the valence band, whereas 
electrons may be quantum-mechanically reflected or transmitted. Electrons entering the 
semiconductor from the plasma may be emitted back to it due to collisions with phonons inside the 
solid, which also cause energy relaxation in its valence and conduction band. To drive a 
current through the setup, a bias voltage is applied between the bulk of the argon plasma, which we 
provide with a prescribed plasma density, and the Ohmic contact used to collect the current.  

From the distribution functions for the charge carriers on both sides of the interface, 
we calculated the current-voltage characteristic. Due to the quantum-mechanical reflection at the
plasma-solid interface and the collisions inside the solid, the characteristic differs significantly
from the one obtained for an interface which absorbs electrons from the plasma perfectly and keeps 
them in the solid forever. Hence, the electron microphysics of the semiconductor affects the electric
properties of the interface and should thus be considered by its theoretical description as well as 
its experimental analysis.

We focused in this work on the implementation of a numerical scheme for the selfconsistent 
calculation of the distribution functions and potential profiles building up at the flux-carrying
plasma-solid interface. For that purpose, we kept the argon plasma collisionless and allowed 
electrons and holes inside the germanium layer to scatter only on phonons. Particle 
nonconserving collisions are not included. Within the simplified model, we find flux-carrying 
conduction band electrons to remain at the high energies set by the injection process. Due to 
the boundary and matching conditions, the surplus electrons inside the semiconductor are thus 
rather hot and not in thermal equilibrium with the intrinsic electrons. The same holds for injected 
holes. With increasing layer thickness, we find secondary electron emission, encoded in the emitted 
electron flux, to increase because a surplus electron is then more likely to suffer collisions with 
phonons, which may bring it back to the plasma, if it also successfully traverses the potential step 
at the plasma-solid interface. The selfconsistent distribution functions enable us, moreover, to 
visualize how the ambipolar gaseous charge transport in the argon plasma merges with the solid-bound 
ambipolar charge transport inside the germanium layer. 

Albeit the description of the argon plasma is also somewhat crude, containing an unspecified 
source and a collisionless sheath made consistent with the germanium layer due to an equally 
unspecified presheath, the idealized electronic structure of the germanium layer is more 
limiting. A quantitative modeling of the biased plasma-solid interface has to be based on a 
realistic band structure of the plasma-facing semiconductor. It should contain, over the energy 
range set by the injection processes, the electronic structure of the surface, entering the 
calculation of the electron reflectivity and the modeling of the hole source function, as well 
as the electronic structure of the bulk, which enters the collision integrals of the Boltzmann 
equations and determines the velocity of the charge carriers. Of particular importance are Bragg 
gaps preventing the transmission of electrons across the interface and surface states trapping 
electrons and/or holes close to it. Since the injection of electrons and holes occurs at rather 
high energies, impact ionization, creating electron-hole pairs across the band gap, and its inverse, 
the recombination of electron-hole pairs have to be moreover also included in a modeling which 
attempts to provide more information than the insight that current-carrying conduction band 
electrons and valence band holes are rather hot. 
%Further progress in combining gaseous and solid-bound 
%electronics could then be made.

\section*{Acknowledgments}
Support by the Deutsche Forschungsgemeinschaft through project BR-1994/3-1 is greatly acknowledged.

\bibliography{ref}
\end{document}